# Frameless ALOHA with Reliability-Latency Guarantees

Čedomir Stefanović[†], Francisco Lázaro[*], Petar Popovski[†]
[†]Department of Electronic Systems, Aalborg University
Aalborg, Denmark. Email: {cs,petarp}@es.aau.dk
[*]Institute of Communications and Navigation of DLR (German Aerospace Center),
Wessling, Germany. Email: Francisco.LazaroBlasco@dlr.de

*Abstract*—One of the novelties brought by 5G is that wireless system design has increasingly turned its focus on guaranteeing reliability and latency. This shifts the design objective of random access protocols from throughput optimization towards constraints based on reliability and latency. For this purpose, we use frameless ALOHA, which relies on successive interference cancellation (SIC), and derive its exact finite-length analysis of the statistics of the unresolved users (reliability) as a function of the contention period length (latency). The presented analysis can be used to derive the reliability-latency guarantees. We also optimize the scheme parameters in order to maximize the reliability within a given latency. Our approach represents an important step towards the general area of design and analysis of access protocols with reliability-latency guarantees.

## I. INTRODUCTION

In a number of emerging wireless technologies, there is an increased focus on offering low-latency services with high reliability[1] guarantees [1]. In this respect, perhaps the biggest novelty that will be brought by 5G is the mode of Ultra-Reliable Low Latency Communication (URLLC) [2]. An important contributor to latency in wireless networking is the access protocol, through which multiple uncoordinated devices attempt to connect to the same access point (AP). The design of random access protocols has traditionally been focused on optimizing the average throughput. The focus on reliability-latency guarantees changes the design problem of an access protocol, which sets the main motivation for this article.

Random access protocols in mobile cellular networks have been based so far on classical slotted ALOHA (SA). Specifically, the well-known result states that the average throughput of SA under the collision channel model is $1/e \approx 0.37 \, \text{packet/slot}$. Another line of works investigated the expected delay performance of SA and its relation to the protocol stability [3]. On the whole, it is easy to draw a conclusion that the classic SA is inadequate to support URLLC services. At the moment of writing, 3GPP has so far considered variants of semi-persistent scheduling as the solution for the uplink URLLC services [4], [5], but this approach is suitable only for devices with periodic traffic patterns, as it otherwise exhibits very low efficiency. In this context, random access protocols may indeed be required to play a role in the execution of URLLC, as a building block of *grant-free* solutions.

Recently, the design space of SA has been significantly expanded by the introduction of successive interference cancellation (SIC) [6]. Following this work, in [7] it was shown that SIC-based reception for the collision channel model in the protocol variant where the users contend with replicas of their packets is analogous to the iterative belief-propagation decoding of erasure-correcting codes, motivating the use of the theory and tools of codes-on-graphs to design and analyze SA schemes with SIC. It was demonstrated that such approach asymptotically achieves expected throughputs that are close to $1 \, \text{packet/slot}$ [7]. This is the ultimate upper bound for the collision channel model and this result has inspired a number of works dealing with application and adaptation of various erasure-correction coding schemes into the SA framework [8].

In general, SA schemes with SIC were analytically treated in terms of the asymptotic packet-loss rate and the asymptotic throughput performance, through the use of density evolution technique [9]. In contrast, the finite-length performance was investigated via simulations or approximate techniques that are appropriate in the error-floor region [10]. A notable exception can be found in recent works [11], [12], where the expected finite-length packet-loss rate and throughput were analytically derived for frameless ALOHA [13], for the collision channel model and single- and multiple-packet reception, respectively.

In this paper, we focus on frameless ALOHA, a variant of SA with SIC where the slots in which the users contend for access are successively "added" on the wireless medium until the target performance has been reached. So far, the performance parameter that drove the optimization of frameless ALOHA was the expected throughput [11]–[14]. Nevertheless, frameless ALOHA has the inherent potential to embed reliability-latency guarantees as its performance goal, which is the main driver of the work presented in the paper. Specifically, in the paper we extend the results from [11] and analytically characterize the finite-length reliability-latency performance of frameless ALOHA. An original contribution of the paper is the introduction of multiple slot classes, where each class is characterized with a corresponding slot-access probability. The output of the analysis is the probability mass function (pmf) of the number of unresolved users[2] for given number of contending users and given number of slots in a class. Building on this result, we optimize the protocol such that the probability that at least a predefined fraction of users becomes resolved in the contention period of a given length is maximized. For a given target latency, expressed as the number of slots, we show that very high levels of reliability can be guaranteed through an adequate choice of the number of slot classes, number of slots in a class, and the corresponding slot-access probabilities. To the best of our knowledge, this is the first work that produces this type of finite-length characterization for any SA scheme with SIC. Moreover, such characterization is the key constituent for the evaluation of the protocol performance in a comprehensive model that involves packet arrivals, backlog, and retransmission policy.

---

[1]The reliability is here defined as the probability that a packet is successfully received at the access point within a predefined time (i.e., latency) [1].

[2]A user is resolved when its packet is successfully decoded/received.

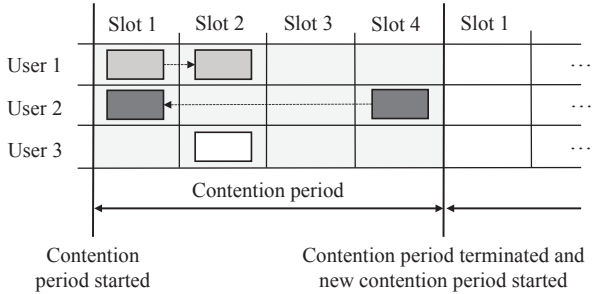

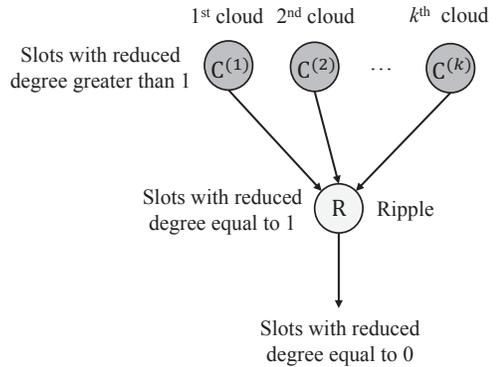

Fig. 1. An example of contention in Frameless ALOHA. All three users randomly and independently decide on a slot basis whether to transmit or not. Slot 1 and slot 2 are collision slots; the colliding transmissions can not be decoded and the AP stores the slots (i.e., the signals observed in them) for later use. Slot 4 is a singleton slot and the AP decodes a replica of the packet of user 2 from it. The AP also learns that a replica of packet of user 2 occurred in slot 1, and removes (cancels) it from the stored signal. Slot 1 now becomes singleton and a replica of the packet of user 1 becomes decoded. In the same manner, the successive process of replica removal and decoding of a new packet replica occurs in slot 2. As all three users have become resolved, the AP terminates the contention period after slot 4, and starts a new one.

Fig. 2. The clouds and the ripple (variables denote the cardinalities).

The organization of the rest of the text is as follows. Section II provides a brief overview of frameless ALOHA and describes the system model. Section III presents the finite-length analysis whose output is the pmf of the number of unresolved users for the given number of slots. Section IV deals with the protocol optimization in order to maximize reliability for the target latency. Section V concludes the paper.

## II. BACKGROUND AND SYSTEM MODEL

### A. Background: Frameless ALOHA

Frameless ALOHA [13] is a variant of SA with SIC that is inspired by rateless coding framework [15]. The time in frameless ALOHA is divided into equal-length slots. The slots are organized in contention periods, where the length of a contention period (in terms of the number of slots) is not a-priori defined. In each slot of the contention period, every contending user independently decides on the slot basis whether or not to transmit a replica of its packet, using a predefined slot-access probability. It is assumed that each replica embeds the information about the slots in which the other related replicas are placed (e.g., via the seed of a random number generator included in the packet header, from which the position of all previous and potential subsequent replicas is extracted). The AP processes slots of the contention period sequentially, decoding and removing replicas. When a predefined criterion becomes satisfied. e.g., the target throughput is reached and/or a predefined fraction of users becomes resolved [14], the AP terminates the current contention period and starts the new one. The start/termination of a contention period can be done via means of beacon signal transmitted by the AP. An example of frameless ALOHA is depicted in Fig. 1.

### B. System Model

We assume that there are $n$ users, contending for the access to a single AP. We focus on a single contention period lasting $m$ slots, where $m$ is not a-priori fixed value, and with $k$ different slot classes. Out of $m$ slots, exactly $m^{(1)}, m^{(2)}, ... m^{(k)}$ belong to slot class $1, 2, ..., k$. In all slots of class $h$, the users use the same slot access probability $p_a^{(h)}$, given by $p_a^{(h)} = \frac{\beta^{(h)}}{n}$. It is easy to verify that $\beta^{(h)}$ is the mean number of users that transmitted in a slot of class $h$, and, thus, the mean number of transmissions contained in the slot.

The contention is performed on the collision channel, whose properties are: (i) singleton slots, i.e., slots containing a single transmission, are decodable with probability 1, and (ii) collision slots, i.e., slots containing two or more transmissions, are not decodable with probability 1. The interference cancellation is assumed to be perfect, i.e., the removal of replicas from the slots leaves no residual transmission power.[3]

In order to model the reception algorithm, i.e., the successive process of decoding and replica removal, we introduce the following definitions:

**Definition 1** (Initial slot degree). *The initial slot degree is the number of transmissions originally occurring in the slot.*

**Definition 2** (Reduced slot degree). *The reduced slot degree is the current number of transmissions in the slot, over the iterations of the reception algorithm.*

**Definition 3** (Ripple). *The ripple is the set of slots of reduced degree 1, and it is denoted it by $\mathscr{R}$.*

The cardinality of the ripple is denoted by r and its associated random variable as R.

**Definition 4** ($h$-th cloud). *The $h$-th cloud is the set of slots of class $h$ with reduced degree $d > 1$, and it is denoted by $\mathscr{C}^{(h)}$.*

---
[3]This assumption is reasonable for practical interference cancellation methods and moderate to high signal-to-noise ratios [7].

The cardinality of the $h$-th cloud is denoted by $c^{(h)}$ and the associated random variable as $C^{(h)}$.

Upon reception, the reduced degree of a slot is equal to its initial degree. If its reduced degree is $d > 1$, the slot of class $h$ is placed in the $h$-th cloud. Each time the slot is affected by the interference cancellation (i.e., a packet replica contained in it becomes removed), its reduced degree is lowered by 1. If its reduced degree $d$ is/becomes 1, the slot is placed in the ripple, when the transmission (i.e., packet replica) contained in it becomes decoded. After decoding, the slots in the ripple are of no further use and their reduced degree is set to 0. This is depicted in Fig. 2. The reception algorithm stops either when all users are decoded, or when the ripple becomes empty.

In the context of the example in Fig. 1, and assuming that there is a single slot class, slots 1 and 2 are in initially in the cloud and slot 4 in the ripple. After fist round of decoding and replica removal, slot 4 leaves the ripple, while slot 1 leaves the cloud and enters the ripple. In the next round of decoding and replica removal, slot 1 leaves the ripple, while the slot 2 leaves the cloud and enters the ripple. The reception algorithm stops when slot 2 leaves the ripple.

Finally, we also introduce slot degree distribution $\mathbf{\Omega}^{(h)} = \left\{\Omega_1^{(h)}, \Omega_2^{(h)}, ..., \Omega_n^{(h)}\right\}$ for the slot class $h = 1, 2, ..., k$, where $\Omega_j^{(h)}$ is equal to the probability that a slot from class $h$ has the initial degree equal to $j$. It is straightforward to verify that $\Omega_j^{(h)}$, $j = 1, 2, ..., n$, is given by

$$\Omega_j^{(h)} = \binom{n}{j} \left(p_a^{(h)}\right)^j \left(1 - p_a^{(h)}\right)^{n-j}$$
$$= \binom{n}{j} \left(\frac{\beta^{(h)}}{n}\right)^j \left(1 - \frac{\beta^{(h)}}{n}\right)^{n-j}.$$

The probability that a slot of class $h$ initially contains no transmission at all is $\Omega_0^{(h)}$, the probability that it initially belongs to the ripple is $\Omega_1^{(h)}$, and the probability that it initially belongs to the $h$-th cloud is $(1 - \Omega_0^{(h)} - \Omega_1^{(h)})$.

## III. FINITE-LENGTH ANALYSIS

For the sake of analysis we shall assume that the receiver works iteratively. If the ripple is empty, the receiver simply stops. Otherwise, it carries out the following steps:
- Selects at random one of the slots in the ripple;
- Resolves the user active in that slot (decodes its packet);
- Cancels the interference contributed by the resolved user from all other slots in which its packet replicas was transmitted. This may cause some slots to leave the cloud and enter the ripple. Furthermore, some slots from the ripple may become degree zero and leave the ripple. These last slots correspond to slots in the ripple in which the resolved user was active.

Thus, in each iteration, reception algorithm either fails, or exactly one user gets resolved. These assumptions are made to ease the analysis and have no impact on the performance.

Following the approach in [11], [16], [17], the iterative reception of frameless ALOHA with $k$ different slot classes is represented as a finite state machine with state

$$S_u := (C_u^{(1)}, C_u^{(2)}, \cdots, C_u^{(k)}, R_u)$$

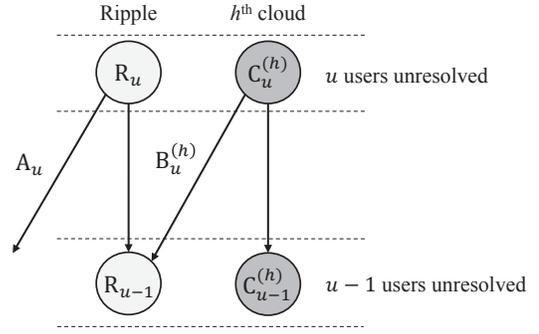

Fig. 3. Evolution of the ripple and the $h$-th cloud through decoding and removal of replicas. Resolution of a single user, i.e., decoding a replica of its packet and removal of the other replicas, causes $A_u$ slots to leave the ripple and $B_u^{(h)}$ slots to leave the $h$-th cloud and enter the ripple.

i.e., the state comprises the cardinalities of the first to $k$-th cloud and the ripple at the reception step in which $u$ users are unresolved. Each iteration of the reception algorithm corresponds to a state transition. The following proposition establishes a recursion used to determine the state distribution.

**Proposition 1.** *Given that its state is $S_u = (c_u^{(1)}, c_u^{(2)}, \cdots, c_u^{(k)}, r_u)$, when $u$ users are unresolved and $r_u > 0$ (i.e., the ripple is not empty), the probability of the receiver being at state $\Pr\{S_{u-1} = s_{u-1}\}$ when $u - 1$ users are unresolved is given by*

$$\Pr\{S_{u-1} = (s_u + w) | S_u = s_u\} = \binom{r_u - 1}{a_u - 1} \left(\frac{1}{u}\right)^{a_u - 1} \times$$
$$\left(1 - \frac{1}{u}\right)^{r_u - a_u} \prod_{h=1}^{h=k} \binom{c_u^{(h)}}{b_u^{(h)}} q_u^{(h) b_u^{(h)}} (1 - q_u^{(h)})^{c_u^{(h)} - b_u^{(h)}}$$

*with*

$$s_u = (c_u^{(1)}, c_u^{(2)}, \cdots, c_u^{(k)}, r_u)$$
$$w = (-b_u^{(1)}, -b_u^{(2)}, \cdots, -b_u^{(k)}, \sum_{h=1}^{k} b_u^{(h)} - a_u)$$

*and*

$$q_u^{(h)} = \frac{\sum_{d=k+1}^{n} \Omega_d^{(h)} \frac{d}{n} \binom{d-1}{k} \frac{\binom{u-1}{k}}{\binom{n-1}{k}} \frac{\binom{n-u}{d-k-1}}{\binom{n-k-1}{d-k-1}}}{1 - \sum_{h=0}^{k} \sum_{d=h}^{n} \Omega_d^{(h)} \frac{\binom{u}{h}\binom{n-u}{d-h}}{\binom{n}{d}}} \quad (1)$$

*for $0 \leq b_u^{(h)} \leq c_u^{(h)}$, $1 \leq a_u \leq r_u$.*

*Proof:* The proof builds up on the proof of Theorem 1 in [11]. We need to consider the variation of the cloud and the ripple cardinalities in the transition from $u$ to $u-1$ unresolved users. In this transition, exactly one user is resolved and all replicas of its packet are removed from the slots in which they appear, which results in some slots leaving the $h$-th cloud, $C_u^{(h)}$ and/or the ripple $R_u$, as depicted in Fig. 3.

Denote by $B_u^{(h)}$ the random variable associated to the number of slots leaving $C_u^{(h)}$, and by $b_u^{(h)}$ its realization. Similarly, denote by $A_u$ the random variable associated to the number of slots leaving the ripple $R_u$, and by $a_u$ its realization. Given the fact that users choose whether to be active or not in every slot independently from other users, every slot is statistically independent from all other slots. Thus, the random variables $B_u^{(h)}$ are independent to each other. From [11] (proof of Theorem 1), we have that the distribution of $B_u^{(h)}$ conditioned to $C_u^{(h)} = c_u^{(h)}$ is binomial with parameters $c_u^{(h)}$ and $q_u^{(h)}$, with $q_u^{(h)}$ given in (1).

The distribution of $A_u$ conditioned to $R_u = r_u$ is [11]

$$\Pr\{A_u = a_u | R_u = r_u\} = \binom{r_u - 1}{a_u - 1} \left(\frac{1}{u}\right)^{a_u - 1} \left(1 - \frac{1}{u}\right)^{r_u - a_u}.$$

The proof is completed by observing that from the definition of $a_u$ and $b_u^{(h)}$ it follows that

$$r_{u-1} = r_u - a_u + \sum_{h=1}^{k} b_u^{(h)} \quad \text{and} \quad c_{u-1}^{(h)} = c_u^{(h)} - b_u^{(h)}.$$

∎

Recall that out of the $m$ slots in the contention period, exactly $m^{(1)}, m^{(2)}, ... m^{(k)}$ belong to slot class $1, 2, ..., k$. We focus on slots of class $h$, of which there are $m^{(h)}$. The initial state distribution corresponds to a multinomial with $m_h$ experiments (slots) and three possible outcomes for each experiment, the slot being in the cloud, the ripple or having degree 0, with respective probabilities $(1 - \Omega_1^{(h)} - \Omega_0^{(h)})$, $\Omega_1^{(h)}$ and $\Omega_0^{(h)}$. Denoting by $R_n^{(h)}$ the random variable associated to the number of slots of class $h$ of reduced degree 1 when all $n$ users are still undecoded, we have

$$\Pr\{(C_n^{(h)} = c_n^{(h)}, R_n^{(h)} = r_n^{(h)}\} = \frac{m^{(h)}!}{c_n^{(h)}! \, r_n^{(h)}! \, (m^{(h)} - c_n^{(h)} - r_n^{(h)})!} \times \left(1 - \Omega_1^{(h)} - \Omega_0^{(h)}\right)^{c_n^{(h)}} \Omega_1^{(h) r_n} \Omega_0^{m^{(h)} - c_n^{(h)} - r_n^{(h)}} \quad (2)$$

for all non-negative $c_n^{(h)}, r_n^{(h)}$ such that $c_n^{(h)} + r_n^{(h)} \leq m^{(h)}$.

Observing that, when all $n$ users are still undecoded, the total number of degree one slots $R_n$ is given by

$$R_n = \sum_{h=1}^{k} R_n^{(h)}$$

we obtain from (2) the initial state distribution of the receiver.

By applying recursively Proposition 1 and initializing as described the finite state machine one obtains the state probabilities. The following theorem is the main result of the paper.

**Theorem 1.** *The probability that exactly $u$ users remain unresolved after a contention period of $m$ slots, $P_u$, is given by*

$$P_u = \sum_{c_u^{(1)}} \sum_{c_u^{(2)}} \cdots \sum_{c_u^{(k)}} \Pr\{S_u = (c_u^{(1)}, c_u^{(2)}, \cdots, c_u^{(k)}, 0)\}.$$

*Proof:* The result is evident from the state machine definition. In particular, the user resolution ends at stage $u$

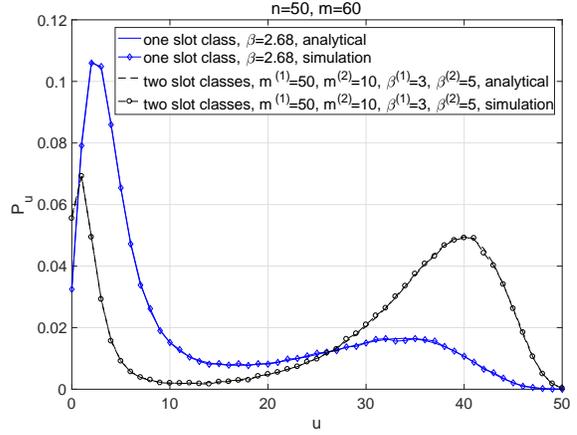

Fig. 4. Examples of probability mass function of the number of undecoded users $u$ for $n = 50$, $m = 60$.

whenever $r_u = 0$ (i.e., whenever the ripple is empty), and this leaves exactly $u$ users unresolved. Thus, we have

$$P_u = \Pr\{R_u = 0\} \quad (3)$$
$$= \sum_{c_u^{(1)}} \sum_{c_u^{(2)}} \cdots \sum_{c_u^{(k)}} \Pr\{S_u = (c_u^{(1)}, c_u^{(2)}, \cdots, c_u^{(k)}, 0)\}$$

where the summations is taken over all possible values of $c_u^{(h)}$, $h = 1, ..., k$. ∎

By applying Theorem 1 for $u = 0, 1, ..., n$, one obtains pmf of the number of unresolved users for the given $n$ and $m$, where the $m$ corresponds to the delay in the number of slots.[4]

As an example, in Fig. 4 we show the pmf of the number of undecoded users $u$, i.e., $P_u$ for $u = 1, ..., n$, when $n = 50$ and $m = 60$, for (i) one slot class with mean initial slot degree $\beta = 2.68$, and (ii) two slot classes, with $m^{(1)} = 50$ slots in the first class, $m^{(2)} = 10$ slots in the second class, with mean initial degrees $\beta^{(1)} = 3$ and $\beta^{(2)} = 5$, respectively.[5] The figure shows analytical results according to Theorem 1 and the outcome of Monte Carlo simulations. We see that the match is tight (100000 contention periods were simulated).

Finally, for the sake of completeness, we also provide result on the the expected packet error rate, i.e., the probability that a user is not resolved, which is derived from (3) as

$$P = \sum_{u=1}^{n} \sum_{c_u} \frac{u}{n} P_u$$
$$= \sum_{u=1}^{n} \sum_{c_u^{(1)}} \sum_{c_u^{(2)}} \cdots \sum_{c_u^{(k)}} \frac{u}{n} \Pr\{S_u = (c_u^{(1)}, c_u^{(2)}, \cdots, c_u^{(k)}, 0)\}$$
(4)

while the expected throughput is simply

$$T = \frac{n(1-P)}{m}.$$

---

[4]Note that $P_u$ implicitly depends on the initial state distribution that is obtained through (2), while (2) depends on the number of slots in a class $m^{(h)}$, $h = 1, 2, ..., k$, and thereby on the total number of slots $m$.

[5]Recall that slot access probability of a class $h$ is $p_a^{(h)} = \beta^{(h)}/n$.

TABLE I
OPTIMAL PARAMETERS FOR FRAMELESS ALOHA THAT MAXIMIZE $\mathsf{F}_t$
FOR $t \in \{48, 50\}$, WHEN $n = 50$ AND $m = 100$, AND $k \in \{1, 2, 3\}$.

| | $t$ | 48 | 50 |
|---|---|---|---|
| $k = 1$ (one slot class) | $\beta$ | 2.9 | 3.33 |
| | $\mathsf{F}_t^{\max}$ | 0.9985 | 0.9934 |
| $k = 2$ (two slot classes) | $m^{(1)}$ | 88 | 86 |
| | $\beta^{(1)}$ | 2.4 | 2.53 |
| | $m^{(2)}$ | 12 | 14 |
| | $\beta^{(2)}$ | 12.94 | 22.08 |
| | $\mathsf{F}_t^{\max}$ | 0.99965 | 0.9986 |
| $k = 3$ (three slot classes) | $m^{(1)}$ | 45 | 88 |
| | $\beta^{(1)}$ | 2.37 | 2.51 |
| | $m^{(2)}$ | 45 | 11 |
| | $\beta^{(2)}$ | 2.47 | 17.39 |
| | $m^{(3)}$ | 10 | 1 |
| | $\beta^{(3)}$ | 12.71 | 50 |
| | $\mathsf{F}_t^{\max}$ | 0.999783 | 0.99917 |

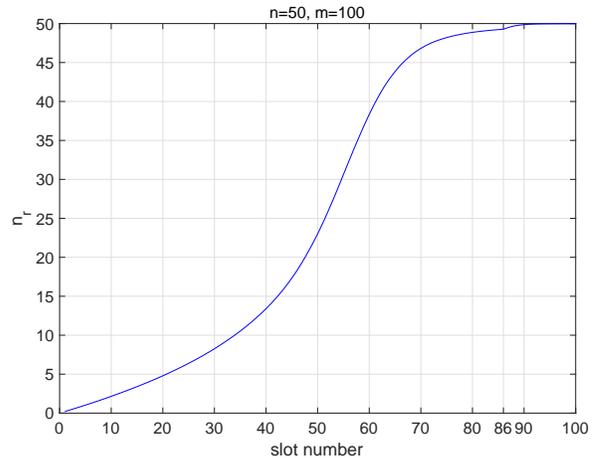

Fig. 5. Evolution of the mean number of resolved users $n_r$ as the slots of the contention period progress, for $n = 50$, $m = 100$ and the scheme with two slot classes from Table I optimized for $t = 50$.

## IV. OPTIMIZATION

The results derived in the previous section can be used to perform optimizaton of frameless ALOHA in different ways. In this paper, we focus on maximization of the probability that at least $t$ out of $n$ contending users are resolved after a contention period of $m$ slots, where $m$ corresponds to the allowed latency. We denote this probability as $\mathsf{F}_t$, and it is computed as

$$\mathsf{F}_t = \sum_{u=0}^{n-t} \mathsf{P}_u = 1 - \sum_{u=n-t+1}^{n} \mathsf{P}_u.$$

Obviously, $\mathsf{F}_t$ corresponds to the cumulative mass function of the pmf $\mathsf{P}_u$ derived in the previous section, evaluated in the point $n - t$, where $n - t$ is the number of unresolved users when $t$ users are resolved.

The optimization parameters are number of slot classes $k$, number of slots in the class $h$, i.e., $m^{(h)}$, and their mean initial degrees $\beta^{(h)}$, for $h = 1, ..., k$, with the constraint that $\sum_{h=1}^{k} m^{(h)} = m$. Since $P_u$ and, thus, $F_t$, implicitly depend on the optimization parameters, the straightforward approach is to employ exhaustive search. However, this quickly becomes impractical. Instead, we performed a joint optimization over all parameters using Nelder-Mead simplex search method in Matlab [18]. For fixed $k$, $n$ and $m$, we perform several optimizaton runs, each from a different starting point, since a run may converge to local minima/maxima, and take the best solution.

We demonstrate the optimization results for $n = 50$, $m = 100$ and for $t \in \{48, 50\}$, i.e., the target is to maximize the probability that at least 96 % and 100 % of users becomes resolved after 100 slots, respectively. Table I shows the optimal value of parameters and the maximum values of $F_t$, when the number of classes $k$ is varied from 1 to 3; note that when $k = 1$, the number of slots in the class is equal to the total number of slots, i.e., equal to 100.

The results demonstrate that increasing number of classes pushes the performance in terms of number of "nines". However, identification of the optimal number of classes that, for the given $t$, $n$ and $m$, provide the overall maximum of $\mathsf{F}_t$ is an open question. Another insight is that, for the given number of classes $k$, the more slots in a class, the lower the mean initial degree of the slots in the class. This is somewhat reminiscent of the optimal degree distributions for LT codes [19], which are a category of rateless codes, and where, in general, the high degrees are less probable then low degrees. The purpose of the slots with high mean initial degrees (i.e., high slot-access probabilities) is to increase the probability that a user actually transmits in the contention period, which is the necessary precondition for the user to become resolved. This is demonstrated in Fig. 5, which shows the evolution of the mean number of resolved users, denoted by $n_r$, for the scheme with two slot classes from Table I optimized for $t = 50$, where it is assumed that initial batch of 86 slots belongs to class 1, and the final 14 slots belong to class 2.[6] Note that $n_r$ is computed using (4) as $n_r = n(1-\mathsf{P})$, where $\mathsf{P}$ is computed for each slot of the contention period. A closer inspection reveals that, after $86^{\text{th}}$ slot, there is a boost in $n_r$, which is due to the increase in $\beta^{(2)}$ with respect to $\beta^{(1)}$.

On the other hand, high initial slot degrees may be problematic from the perspective of the actual reception of a composite signal in a slot and when the interference cancelation is non-ideal. A potential approach to alleviate this issue in systems where downlink channel is available is the one where AP acknowledges the resolved users, which then abstain from transmitting in the rest of the contention period and thus effectively decrease the initial slot degrees.

---

[6]Here we remark that slots from different classes do not necessarily have to appear in continuous batches if one is interested only at the performance at the end of the contention period, which is the case considered in this paper. In particular, the performance at the end of the contention period will be the same regardless of the organization of the slots in the contention period, if the "stipulated" number of slots from each class is met.

*Adaptive Frameless ALOHA*

The optimization performed in the previous text is static, in the sense that all the parameters are defined a-priori. On the other hand, in the systems with a "balanced" uplink and downlink, like in mobile cellular systems, the downlink can be used to dynamically, i.e., adaptively drive the contention process by periodically informing the users of the optimal slot-access probability that should be subsequently used. The optimal slot access probability can be derived on the basis on the current state of the ripple and the clouds, e.g., using Markov decision processes framework. The detailed analysis and derivation of the update rules is beyond the paper scope. Nevertheless, we demonstrate the potential of the adaptive approach using a heuristic rule in which slot-access probability $p_a$, i.e., the corresponding $\beta$, is updated after every slot as

$$\beta = \frac{n}{u}\left(1 + (\beta^* - 1)\frac{m - (n - u) - c_u}{m}\right)$$

where $n$ is the number of users, $u$ is the current number of unresolved users, $m$ is the (total) number of slots, $c_u$ is the current cardinality of the cloud (here the cloud comprises all slots with reduced degree greater than 1), and $\beta^*$ is the mean initial degree that maximizes the peak throughput in the static optimization for the single slot class, see [11]. The motivation behind this heuristic is the following: (i) if no users have been decoded, i.e., $u = n$, and the cloud is empty, i.e., $c_u = 0$, $\beta$ is set to $\beta^*$, (ii) if there is a large number of slots in the cloud, $\beta$ takes a value close to $n/u$, which makes a new slot belong to the ripple with high probability, (iii) otherwise, $\beta$ is set to a value between $n/u$ and $\beta^*$. For $n = 50$, $m = 100$, $\beta^* = 2.47$, and $t = 50$, it can be shown via means of simulation that this strategy achieves $F_t = 0.9999975$. I.e., all users are resolved with reliability of more than 5 nines after 100 slots.

## V. Conclusions and Discussion

We presented finite-length analysis of frameless ALOHA for collision channel model, which could be used to assess the reliability of the scheme for the given contention period length. We also performed optimization of the scheme, such that probability that at least the target fraction of users becomes resolved at the end of the contention period. The obtained results are promising, showing that very high probabilities of user resolution can be reached. To the best of our knowledge, this the first work dealing with the exact derivation of the reliability-latency results for any SA scheme with SIC.

The assumed framework in which the latency (i.e., the contention period length) is fixed may seem to oppose the frameless nature of the protocol. However, reliability guarantees naturally involve latency deadlines, and the design goal was to tune the distributions to drive the protocol in a way that, after a given deadline, a statistical guarantee can be offered.

The presented work can be extended in several ways. As already hinted, further investigations can be made in terms of the identification of the optimal number of slot classes for given $n$, $m$ and $t$, as well as of the achievable bound on $F_t$. Also, the results presented in Fig. 4 show that the obtained pmfs are bimodal; in this respect, investigating the shape of the pmf as function of the scheme parameters, as well as its explicit modeling seem to pose interesting problems. Another potential extension is an optimization of the scheme not only with respect to the performance at the end of the contention period, but also taking into account its intermediate performance. Such approach could be used when one is interested in a balanced latency profile; in terms of Fig. 5, this would imply a more "linear" increase of the number of resolved users. Another, similar case would be the one in which the intermediate performance guarantees are of interest. The complexity of the used optimization approach quickly increases with the number of slot classes $k$. A part of our on-going work is devoted to low-complexity approximations. Finally, the results show that assessment and optimization of the dynamic, adaptive version of the scheme are also worth considering. These tasks are also part of our ongoing work.

## Acknowledgment

The work of Č. Stefanović and P. Popovski was supported in part by the European Research Council (ERC Consolidator Grant Nr. 648382 WILLOW) in the Horizon 2020 Program.